# ATMOSPHERIC CONSEQUENCES OF COSMIC RAY VARIABILITY IN THE EXTRAGALACTIC SHOCK MODEL: II Revised ionization levels and their consequences

Adrian L. Melott[1], Dimitra Atri[1], Brian C. Thomas[2], Mikhail V. Medvedev[1], Graham W. Wilson[1], and Michael J. Murray[1]

**ABSTRACT**
It has been suggested that galactic shock asymmetry induced by our galaxy's infall toward the Virgo Cluster may be a source of periodicity in cosmic ray exposure as the solar system oscillates perpendicular to the galactic plane, thereby inducing an observed terrestrial periodicity in biodiversity. There are a number of plausible mechanisms by which cosmic rays might affect terrestrial biodiversity. Here we investigate one of these mechanisms, the consequent ionization and dissociation in the atmosphere, resulting in changes in atmospheric chemistry which culminate in the depletion of ozone and a resulting increase in the dangerous solar UVB flux on the ground. We use a heuristic model of the cosmic ray intensity enhancement originally suggested by Medvedev & Melott (2007) to compute steady-state atmospheric effects. This paper is a re-examination of an issue we have studied before (Melott et al. 2008) with a simplified approximation for the distribution of incidence angles. The new results are based on an improved ionization background computation averaged over a massive ensemble (about $7 \times 10^5$) shower simulations at various energies and incidence angles. We adopt a range with a minimal model and a fit to full exposure to the postulated extragalactic background. The atmospheric effects are greater than they were with our earlier, simplified ionization model. At the lower end of the intensity range, we find that the effects are too small to be of serious consequence. At the upper end of this range, ~6 % global average loss of ozone column density exceeds that currently experienced due to anthropogenic effects such as accumulated chlorofluorocarbons. We discuss some of the possible effects. The intensity of the atmospheric effects is less than those of a nearby supernova or galactic gamma-ray burst, but the duration of the effects would be about $10^6$ times longer. Present UVB enhancement from current ozone depletion ~3% is a documented stress on the biosphere, but a depletion of the magnitude found at the upper end of our range would approximately double the global average UVB flux. We conclude that for estimates at the upper end of the reasonable range of the cosmic ray variability over geologic time, the mechanism of atmospheric ozone depletion may provide a major biological stress, which could easily bring about major loss of biodiversity. It is possible that future high energy astrophysical observations will resolve the question of whether such depletion is likely.

[1] University of Kansas, Department of Physics and Astronomy, 1251 Wescoe Dr. # 1082, Lawrence, KS 66045-7582; melott@ku.edu, dimitra@ku.edu , medvedev@ku.edu, gwwilson@ku.edu, mjmurray@ku.edu
[2] Washburn University, Department of Physics and Astronomy, Topeka, KS 66621; brian.thomas@washburn.edu

# 1. Introduction

A 62 ± 3 My periodicity in fossil biodiversity that goes back more than 500 My has been found in one paleontological database at better than 99% confidence [*Rohde and Muller* 2005, hereafter RM; *Cornette* 2007; *Lieberman and Melott* 2007]. More recently, it has also been found [*Melott* 2008; *Melott and Bambach* 2010a] at the same period and phase in two other independent databases. *Medvedev and Melott* [2007; hereafter MM] proposed a mechanism for this periodicity based on the motion of the Sun and Solar System perpendicular to the galactic plane. Although there is considerable uncertainty in the global mass density of the galaxy, the motion is quasiperiodic (due to nonuniformities) at about 63.6 My [*Gies and Helsel* 2005; *Binney* 2006], equal within (systematic) uncertainties to that of the biodiversity periodicity, and in phase with it. Departures from axial symmetry due to spiral arms introduce scatter in such solutions of only a few My over the last 500 My, as shown by *Svensmark* [2006]. This is comparable to the size of the bandwidth in the RM fit to the principal periodicity. Thus, any effect lowering biodiversity upon excursion to Galactic north would have the right sense.

MM proposed that there are heightened cosmic rays (CR) of energies ~TeV and above from a galactic termination shock and possibly a bow shock as well, associated with the Galaxy's well-known infall toward the Virgo Cluster. The solar system is shielded from these cosmic rays when inside or south of the galactic disk by the turbulent magnetic field of the galaxy, but there is substantially increased exposure at galactic north. In distinction to sources such as gamma-ray bursts or solar flares studied in our previous work on astrophysical ionizing radiation effects, this excess irradiation would persist for ~$10^7$ y of each cycle. In the past [*Melott* et al. 2008] we presented results approximating an isotropic flux by an "average" zenith angle. Here we expand that with a massive Monte Carlo simulation including a wide range incidence angles, with more runs for each incidence angle.

One fairly obvious possibility for impact on biodiversity is direct effects of radiation. Although most of the energy of cosmic rays is deposited in the atmosphere, some reaches the ground particularly in the form of muons (μ) which comprises about 85% of the equivalent biological dose of CR in the USA; in turn CR products constitute about one-third of the annual dose in the USA [*Alpen* 1998]. Increased exposure with altitude is the presumed cause of increased cancer rates among airline personnel [e.g. *Pukkala* et al. 1995]. Cancer rates correlate strongly with cosmogenic $^{10}$Be [primarily formed in CR interactions with the atmosphere] as well as with latitude, in an apparent effect on the germ cells of one's parents [*Juckett* 2007, 2009]. Low-energy cosmic ray dose increases toward the poles, due to guiding effects of the Earth's magnetic field. Study of bone cancer in dinosaur fossils shows no evidence for a radiation effect [*Natarajan* et al. 2007]. It is consistent with the data that the causal agent of the 62 My periodicity may act as a background stress, rendering other kinds of events more serious, and transforming them into mass extinctions [Melott and Bambach 2010b].

An additional class of effects is connected to changes in atmospheric chemistry. The triple bond of $N_2$, which comprises about 80% of the atmosphere, is very strong. Organisms have devised a number of strategies for breaking this bond so that nitrogen can become chemically available. CRs or high-energy photons are able to break this bond high in the atmosphere, which leads to formation of a number of compounds which are normally only present at low abundance. This includes NO and $NO_2$, which are referred to as $NO_x$. These compounds persist in the atmosphere for years, being slowly removed primarily by rain or snow precipitation as nitric acid ($HNO_3$), as well as by photolysis. For details see *Thomas* et al. [2005].

There are two main overt chemical effects of potential importance. (1) The most important effect of $NO_x$ appears to be ozone ($O_3$) depletion [*Ruderman* 1974; *Ejzak* et al. 2007, and references therein]. $NO_x$ catalyze the conversion of $O_3$ to ordinary oxygen molecules. Stratospheric $O_3$ absorbs strongly in the UVB (280-315 nm) band, removing about 90% of it. UVB is strongly absorbed by and damages DNA and proteins, and is a known carcinogen and mutagen. Possible large Solar flares, nearby supernovae and galactic gamma-ray bursts are candidates for catastrophic ozone depletion. Disastrous effects are possible at high depletion levels [*Melott and Thomas* 2009]. (2) Secondarily, the brown gas $NO_2$ may have some effects of opacity [*Melott et al.* 2005]. This effect is subdominant to ozone depletion, and is only likely to be serious in extreme cases. We have examined the opacity as described in the next section. We also expect some production of $NO_x$ and (ironically) $O_3$ in the troposphere. These compounds are toxic. This is not an issue in previous studies based on ionizing photons, but these high energy cosmic rays with lower cross sections and more penetrating power due to increased momentum, are such that we expect energy deposition lower in the atmosphere. We have studied these two effects and report our conclusions in this paper.

There is a large and somewhat controversial literature on the association of low-altitude cloud formation with cosmic rays. For a review see *Scherer et al.* [2007]; however, some of the original motivation for this association has evaporated, due to new information on the structure of Galactic spiral arms [*Overholt* et al. 2009]. Clouds may increase albedo, back-scattering solar photons and lowering the global temperature. The most robust conclusion seems to be that CR flux probably forces climate on long timescales, but that recent warming cannot be attributed to reduced cosmic ray flux [*Lockwood and Fröhlich* 2007]. The CRs on which past studies were based are low energy, modulated by the Earth's magnetic field and the solar wind. TeV CRs would not be so modulated. Furthermore, their energy deposition lower in the atmosphere might be even more conducive to low-level cloud nucleation. Lower temperatures from increased albedo, if present, would be expected to lower biodiversity [*Aguirre and Riding* 2005, and references therein].

The purpose of this study is to model the potential atmospheric chemical effects of the increased CR flux associated with the MM hypothesis, and estimate the magnitude of potential biological or climatic consequences based on $NO_x$ production and $O_3$ depletion. Direct radiation effects on organisms and possible cloud nucleation mechanisms are not included in this study.

## 2. Methods

The hypothesis of enhanced cosmic ray flux correlated with excursions of the Sun to Galactic north assumes that there will be substantially lower shielding from the galactic magnetic field at those times. The enhanced flux originates in CRs accelerated at shocks which are closer to the Galaxy on the north side, toward the Virgo Cluster (MM). Only rough estimates are available of the magnitude and functional shape of the enhanced spectrum. We will parameterize the CR flux spectrum as a function of CR energy E and our displacement z from the galactic plane. This parameterization is based on the model described in MM, where the slope beyond the knee (-3.2) is assumed to be the "true" flux in the neighborhood of our Galaxy, suppressed by the magnetic field inside the disc.

$$F(z,E) = C_n \{E_P^{-2.7} \exp[-(E_P/3)^{0.5}] + f(z) E_P^{-3.2} \exp(-E_0/E_P)\} \qquad [1]$$

Where $C_n$ is a normalization constant, $E_P$ is the energy in PeV, $E_0$ is a variable cutoff in the same units, and $f(z)$ parameterizes the enhancement due to solar position. The first term represents the normal, lower-energy contribution by sources inside the galaxy [presumably largely supernova remnants]. When $E_0 = 3$ PeV, $f(z)=1$, the CR spectrum observed locally, where z is small (~8 pc) is reproduced [to within the accuracy we need], and the contributions approximately meet at the knee. This functional form is not particularly valid for E>3 or for $f(z)<1$; this does not concern us because contributions from energies above the knee are negligible, even in the enhanced state, and we will not consider the suppressed state, ie when z<0. Estimates of the range of enhancement consistent with extragalactic backgrounds, pressure, etc. lead us to study 2 cases here: $E_0 = 10^{-3}$ in case 1, and $10^{-4}$ in case 2; and $f(z)$ adjusted so that the subsequent flux at 1 TeV is enhanced by a factor of 4 (case 1) or 25 (case 2). Case 1 is an estimate based on the best estimates of relevant parameters, and Case 2 the maximal value consistent with existing constraints (MM). While far from exhaustive, this will enable us to do a preliminary exploration of parameter space in order to determine whether interestingly large atmospheric effects are possible and worthy of more detailed exploration in the future. In Figure 1 we show the range of CR spectra we investigate in our model, including the unperturbed case. We do not include the present lower-energy background in these figures, because its energy deposition is built into the atmospheric code (described below).

We performed our modeling using the Goddard Space Flight Center two-dimensional (latitude, altitude) time-dependent atmospheric model that has been used extensively to study the effects of solar flares, as well as supernovae and gamma-ray burst effects. We will describe the code only briefly, given accounts elsewhere [see *Thomas et al.*, 2005; *Ejzak et al.*, 2007; *Thomas et al.*, 2007; and references therein]. There are 18 bands of latitude and 58 log pressure bands. The model computes atmospheric constituents in a largely empirical background of solar radiation variations, with photodissociation, and including winds and small scale mixing. It also includes an empirical background of CR source ionization, which includes an 11-year solar modulation cycle, all with a timestep of one day.

Our CRs are of much larger primary energy than normal galactic CR, so we cannot simply turn up the usual background, as was done in supernova studies [*Gehrels et al.* 2003]. CRs of energy 100 GeV and up are likely to form the enhanced population; they will not be particularly affected by the solar wind or geomagnetic field, and due to smaller cross sections and more momentum (penetrating power) will deposit energy lower in the atmosphere. To simulate the air showers we used CORSIKA [e.g. *Djemil et al.* 2005; see http://www-ik.fzk.de/corsika/], a program for detailed simulation of extensive air showers initiated by high energy cosmic ray particles. Our procedure and the openly available table of our results is described elsewhere in more detail [*Melott et al.* 2010]; we summarize it here briefly. The table will later be extended beyond the range needed here, to allow for other possible uses. We use this energy deposition information convolved with our assumed spectra of enchanced CR, considerably extending the range of the NGSFC code and making possible the treatment of energetic particles of arbitrary energy up to PeV as a source of atmospheric ionization.

It is computationally unfeasible at this time to do Monte Carlos over all possible angles of incidence, so we investigated and applied an approximation scheme. We modeled the effect of zenith angle by doing shower ensembles each at zenith angles of 5° (from vertical), 15°…85°, at ten degree intervals over the range of energies noted above. An isotropic flux is modeled by taking contributions from each of these weighted by sin θ. We have done 13500 simulated showers at each of a series of energies separated by 0.1 in log10 of primary energy between 17 GeV and 1 PeV, i.e. at 51 different primary energies. CORSIKA ran 900 times, 100 times each for 9 different zenith angles at a given primary energy. Each time it was run, CORSIKA simulated 15 protons entering the atmosphere, a number set by computational constraints. Therefore, we compiled 13500 particles for each of the 51 energy levels, for a total of about $7 \times 10^5$ particles simulated. The energy was averaged for each of the 900 runs at each longitudinal pressure bin. In Figure 2a we show the fractional energy deposition (excluding nuclear interactions) for the "normal" cosmic ray spectrum $E^{-2.7}$ for two of these zenith angles per interval in log pressure. Pressure is proportional to total column density traversed by the shower, and is approximately linear in altitude. Note that the lateral displacement in the lines, and the location of their maxima, are reasonably approximated by a $(\cos \theta)^{-1}$ factor, the simplest thing one would expect from a column density factor. However, the tails of the distributions do not follow this relation, which could be important for effects in the troposphere.

For each shower, we recorded the fractional energy deposition in each of 1000 bins of 1 g/cm$^2$ column density. We used their mean to construct a lookup table describing the energy deposition for a given primary energy as a function of pressure. We then used this lookup table to construct the total energy deposition of a given spectrum of CR primaries.

The greatest deposition of energy per bin is often in the first or second bin, corresponding to a first interaction very high in the atmosphere. Since log column density is nearly linear with altitude, we plot energy deposition per log column density. In Figure 2b we show this for the lowest, the mid-range, and highest energy primaries relevant here, averaged over incidence angle as described above. This has a trend with primary energy, corresponding to deeper penetration into the atmosphere for higher primary energies, due to more momentum and somewhat lower interaction cross sections.

This gives about 150 to 400 g/cm$^2$, depending on primary energy, as the site of the greatest deposition of energy per unit distance, corresponding to an altitude of about 14-8 km. This can be compared with (a) about 13 km as the altitude of maximal energy deposition density for the normal CR spectrum as implemented in the NGSFC code, strongly biased toward high latitudes, and (b) 22 to 35 km as the peak deposition for keV-MeV photons, depending upon energy [*Ejzak et al*. 2007]. Recall that ozone ($O_3$) depletion is the primary effect of interest here. $O_3$ is predominantly found at altitudes of 10-35 km [e.g. *Harfoot et al*. 2007], with considerable latitude dependence [*Ejzak et al*. 2007]. The results we will present are partially a result of these differing altitude distributions. The differences in penetrations result from a difference in cross sections combined with the fact that the normal dominant CR spectrum up to about a GeV, is strongly guided by the Earth's magnetic field, comes down with a small zenith angle near the poles, and consequently encounters a lower mean column density than an isotropic ensemble.

## 3. Results

Since in the MM hypothesis the solar system resides in a region of excess CR flux for of order $10^7$ y, we take this angle-averaged flux for all points on the globe. Our results will be presented in the steady-state approximation, well after the flux is "turned on", but will still show seasonal and solar cycle fluctuations.

Our primary result is that the changes in atmospheric chemistry are considerably different for our two cases of the EGCR enhancement. In addition, effects are larger than we found in *Melott* et al. [2008] with its simplified approach to angle-averaging. We report here two ways of viewing effects on $O_3$ due to the enhancements: 1) changes in globally averaged $O_3$ column density and 2) changes in profile $O_3$ volume number density. Both are reported as percent differences, comparing point-wise between a run with the EGCR enhancement and a run with only the normal background. From the point of view of ground-level UV, column density is the most important result. We turn on the cosmic ray flux in a background model of the present atmosphere, and run until the transients have been effectively removed, which about a decade. After this point there is no long-term accumulation of $NO_x$. Column density changes are shown in Figures 3a and 3b over 200 months, representative of the steady state. Variations here are due to seasonal and solar cycle changes which affect photolysis reactions that are important for $O_3$. Depletions for case 1 (Figure 3a) range around 0.6%. These are smaller than changes that took place over longer timescales due to changes in atmospheric oxygen concentration and solar luminosity [*Björn and McKenzie* 2007; *Harfoot et al.* 2007]. Furthermore, they are small compared with the level change necessary to induce serious damage to the biosphere (e.g. *Jagger* 1985; *Häder et al.* 2003). We conclude that changes in case 1 are too small to have a major effect on the biosphere, even with assumed $10^7$ year durations.

Ozone depletion level is a nonlinear function of flux, for reasons discussed elsewhere [*Ejzak et al.* 2007]. For case 2 (Figure 3b), column density changes are more significant, ranging around 6%. This is much greater than present ozone depletion due to anthropogenic sources. Present depletion is known to have some role in stress on the biosphere; for a summary see [*Melott and Thomas* 2009]. Earth's orbital resonances will cause rapid variations in the $O_3$ column density at specific seasons, but not in the mean annual column density [*Björn and McKenzie* 2007; *Shaffer and Cerveny* 2004]. There are a variety of geological and biological proxies being developed to monitor past $O_3$ layer levels, but all are in developmental stages at this time. We will present only a summary of the results, arguing that greater detail is not justified in this case. Case 2 appears to be a definite stressor to the biosphere, possibly at a level which might over $10^7$ years bring about considerable extinction of species, especially working together with other events.

Column density changes are the net result of both production and depletion of ozone throughout the altitude profile. Figures 4a and 4b show point-wise percentage changes in profile $O_3$ volume number density at month 252, which corresponds to March and represents a middle value in column density changes for both cases. As seen above, changes are significantly greater for case 2. However, both cases show qualitatively similar effects in profile changes, with some shifts.

In the grayscale version of the images, both depletion and increase are shown in grey, with white for no change. In 4a, nearly all the plot is filled with ozone synthesis, strongest at low altitudes. Depletion is visible near the poles at high altitudes. The asymmetry between poles is primarily a consequence of the season chosen (March). The apparent dominance of synthesis is a result of the fact that the density of ozone is quite low at low altitudes. There is a small net *decrease* in ozone column density averaged over the globe, as shown in Figure 3a.

More definite structure is visible in Case 2, Figure 4b. A border exists between depletion of $O_3$ at higher altitudes and production of $O_3$ at lower altitudes. This border runs from about 15 km at the poles to about 30 km at the Equator. Production at lower altitudes occurs for two primary reasons: 1) Depletion of $O_3$ by $NO_x$ at higher altitudes allows solar UV to penetrate deeper into the atmosphere than normal, which dissociates $O_2$ molecules leading to production of $O_3$. 2) At lower altitudes, $NO_x$ compounds produced by cosmic rays can be photolyzed to produce O which then reacts with $O_2$ to produce $O_3$. We again note that production values here are somewhat visually misleading, appearing relatively large because $O_3$ number densities are normally small at lower altitudes. Figure 5 shows the normal $O_3$ number density profile for this month. We note that in case 2, ground level $O_3$ is increased by up to ~10%, still not a large enough increase to be toxic.

In comparison to our previous studies involving photon events [*Ejzak et al.* 2007, *Thomas et al.* 2005], we see larger production of $O_3$ at low altitudes in the present cases. This is primarily a difference in where energy is deposited in the atmosphere. The photon events typically had a peak energy deposition around 35 km while in these cases the energy deposition peaks at about 10 km. More $NO_x$ is therefore produced at lower altitudes where $O_3$ is normally sparse and hence photochemical production is more efficient.

Our examination of opacity is motivated by the possibility that climate might be cooled by reduced transmission of solar energy to the surface. We find that with the maximal case 2, only a global average of about 0.09% of solar energy, is absorbed by $NO_2$; of course it is much less in case 1. In case 2 the opacity locally reaches levels that would reduce insolation by about 9%, but only in circumpolar regions beginning around the fall equinox. The $NO_2$ quickly dissipates due to photolysis when the Sun returns and will not greatly reduce the melting of polar ice. The amount in Case 2 might bring about some cooling, but not extreme amounts. Based on previous studies [S*olanki and Krivova* 2003; *Foukal et al.* 2006], we estimate that this would induce a cooling of only about 0.1 °C in the maximal case 2, and therefore might have a small effect on biodiversity. We recall that the expected CR flux enhancement is likely to lie between our Case 1 and Case 2.

**4. Discussion**

We have employed a widely-used cosmic ray air shower code to model the energy deposition of cosmic ray primaries in the range 0.1-1000 Tev. We have convolved the resulting distribution function with the spectrum of two cosmic ray energy spectra. The first one is based on the most likely parameters while the 2[nd] has the maximum increase in CR flux consistent with current uncertainties in describing intergalactic shocks. The resulting energy deposition extends the useful energy range of the NASA-Goddard Space Flight Center 2D atmospheric code, to model the response of the atmosphere to the enhanced EGCR flux hypothesized by MM. In doing so, we have performed a first-order evaluation of the plausibility of one of the mechanisms of stress on the biosphere which might result from the enhanced CR flux. We have found that in case 1 the magnitude of the stratospheric $O_3$ depletion and resulting enhanced solar UVB is smaller than that experienced now from anthropogenic causes. In case 2, the effect is greater, with a globally averaged fractional depletion of about 6% in $O_3$ with localized maxima up to 48%.

The reduced $O_3$ allows more UVB, 280-315 nm to reach the surface. The levels we find in the case 2 simulation are larger than those noted from current anthropogenic $O_3$ depletion, currently at a global average of order 3%, where it may have stabilized since the banning of cholorfluorocarbon refrigeration and propellant compounds. The current situation may be used as a kind of template to characterize the effects that might result from enhanced UVB.

UVB has a wide variety of damaging effects on organisms. These include a number of different kinds of damage to DNA; inhibition of growth in seedlings and inhibition of the synthesis of essential organic compounds; photocarcinogenesis and skin edema in animals; damage to the cornea, inhibition of motility in microorganisms [*Madronich* 1993]. Terrestrial plants show a variety of effects, including reduced growth rates and enhanced sensitivity to pollutants and elevated temperature [*Tevini* 1993]. Effects on aquatic ecosystems are severe, affecting motility of organisms and photosynthetic efficiency [e.g. *Boucher and Prezelin* 1996]. There are indications that UVB reduces photosynthesis in Antarctic waters by as much as 25% with current levels of ozone depletion. As much as 20% of the current global primary food production is in the southern oceans [*Häder* 2003]. Field experiments on amphibians [*Blaustein et al.* 1998] show that the hatching success of eggs may be strongly affected, at all altitudes. A meta-analysis of UVB effects [*Bancroft et al.* 2007] found an overall negative effect on both survival and growth across life histories, trophic groups, habitats, and experimental venues. Despite the overall negative effect, there were large variations of intensity: for example, embryos were affected quite strongly. Contrary to the experimental hypothesis, there were no systematic differences such as between fresh water and marine habitats, or with altitude. UVB was found to interact synergistically with other stressors such as contaminants, disease, and extreme thermal events. In fact, UVB acts in a strongly synergistic way with nitrate [*Hatch and Blaustein* 2000], and nitrate would also be enhanced as the $NO_x$ are rained out of the atmosphere [*Melott et al.* 2005]. This, combined with diffusion down from the stratosphere, is the primary removal mechanism after the initial transient bout of photolysis. Elsewhere [*Thomas and Honeyman* 2008], we have examined whether the nitrate deposition following a large ionization event (a 100 kJ m$^{-2}$ GRB) would be great enough to cause enhanced amphibian damage, in combination with elevated UVB levels. We have found that the increase in nitrate concentrations would not be significant in that case. Given that the present study involves similar atmospheric effects it is unlikely that nitrate deposition would have much effect here.

UVB effects strongly contribute to amphibian decline [*Kiesecker et al.* 2001] which is proceeding at over 200 times the background rate [*McCallum* 2007]. Major effects are being seen in plankton, with the present UVB increase [*Davidson* 1998]. On the other hand, one cannot discount the possibility of an evolutionary adaptation to the increased UVB levels, and UVB cannot be shown to be the sole cause of these declines. For more discussion, see [*Melott and Thomas* 2009].

We recall that while atmospheric effects in Case 1 are probably too small to be significant, those in Case 2 are greater than present-day $O_3$ depletion, which is widely acknowledged to be a crisis at least for oceanic surface life and amphibians. It is likely that Case 2, which was constructed as a strongest-case scenario, if the steady state were continued for ~$10^7$ years as indicated by the orbital motion of the Sun in the Galaxy, would place additional stress on biodiversity. Clearly, more work is needed to constrain the range of possible cosmic ray intensities in the galactic halo, including high-energy observations (MM) of possible gamma-ray flux from interaction with neutral hydrogen clouds.

Increased atmospheric ionization may have other effects. Also, increased numbers of high-energy muons and thermal neutrons should reach the ground. In future work, we plan to investigate the potential for climate change due to enhanced cloud cover and the direct biological effects of cosmic-ray secondaries at ground level.

## 5. Acknowledgments

This research was supported at Washburn University and the University of Kansas by NASA grant NNX09AM85G. The extensive computer simulations were supported by the NSF via TeraGrid allocations at the National Center for Supercomputing Applications, Urbana, Illinois, TG-PHY090067T and TG-PHY090108. We thank Tanguy Pierog for assistance in our local implementation of CORSIKA.

## 7. Figures

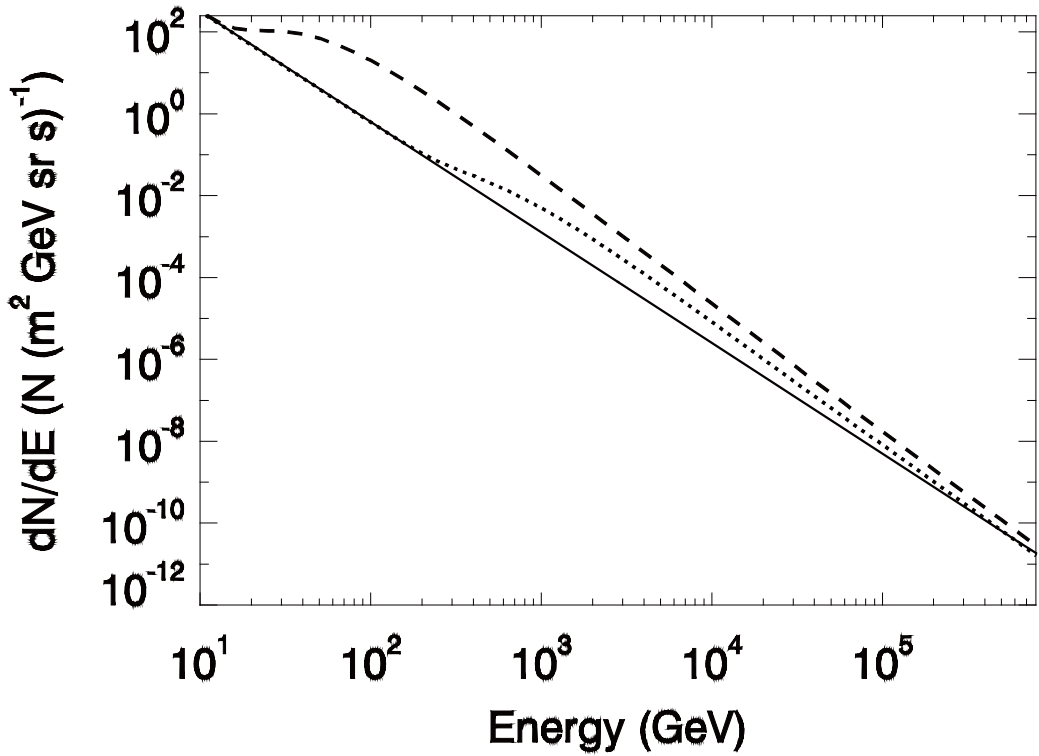

Figure 1: We show schematically the present cosmic ray flux spectrum incident at the top of the Earth's atmosphere (solid line), and our minimum (short dash, called case 1) and maximum (long dash, called case 2) estimates of the spectrum enhanced due to reduced shielding from CRs produced at the galactic bow shock, when the solar system is at 70 pc north of the galactic disk. In this paper we assess atmospheric chemistry changes due to the two cases which bracket the reasonable range of intensity.

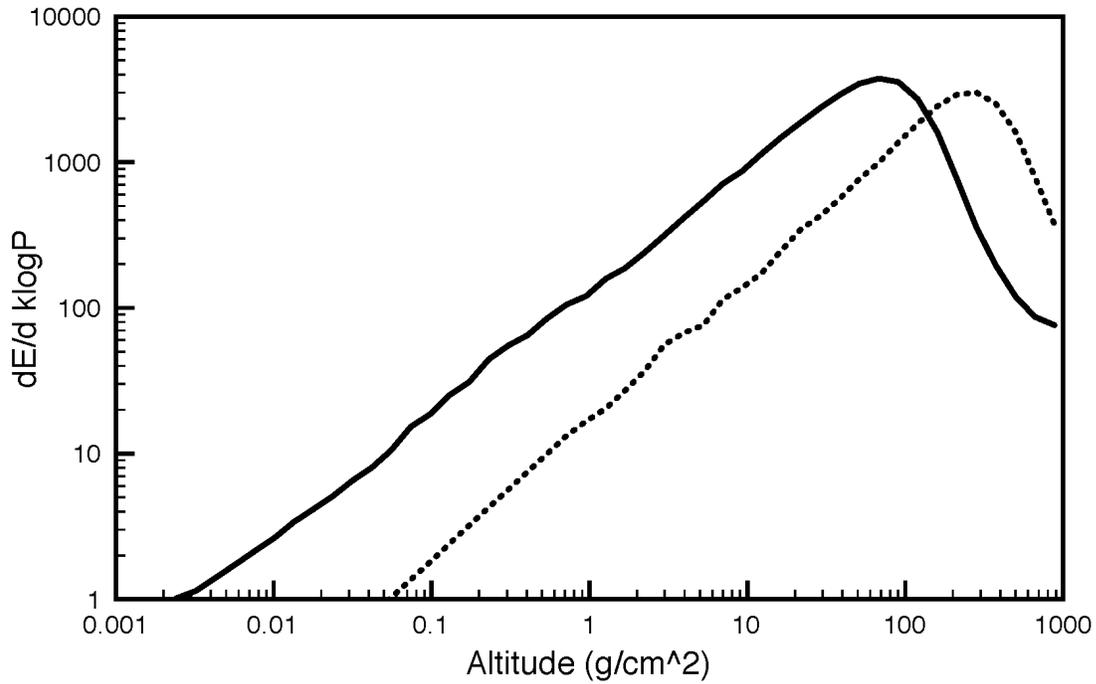

Figure 2a: We show the distribution of atmospheric energy deposition for ensembles of CORSIKA simulations for the cosmic ray spectrum with primaries at 10 GeV to 1 PeV, weighted as $E^{-2.7}$ in analogy with the present observed spectrum at the Earth. This is plotted as fraction of the energy of the primary versus log of the column density (nearly linear with altitude) in bins of 10 g/cm$^2$. The zenith angle is 5° (dotted) and 75° (solid).

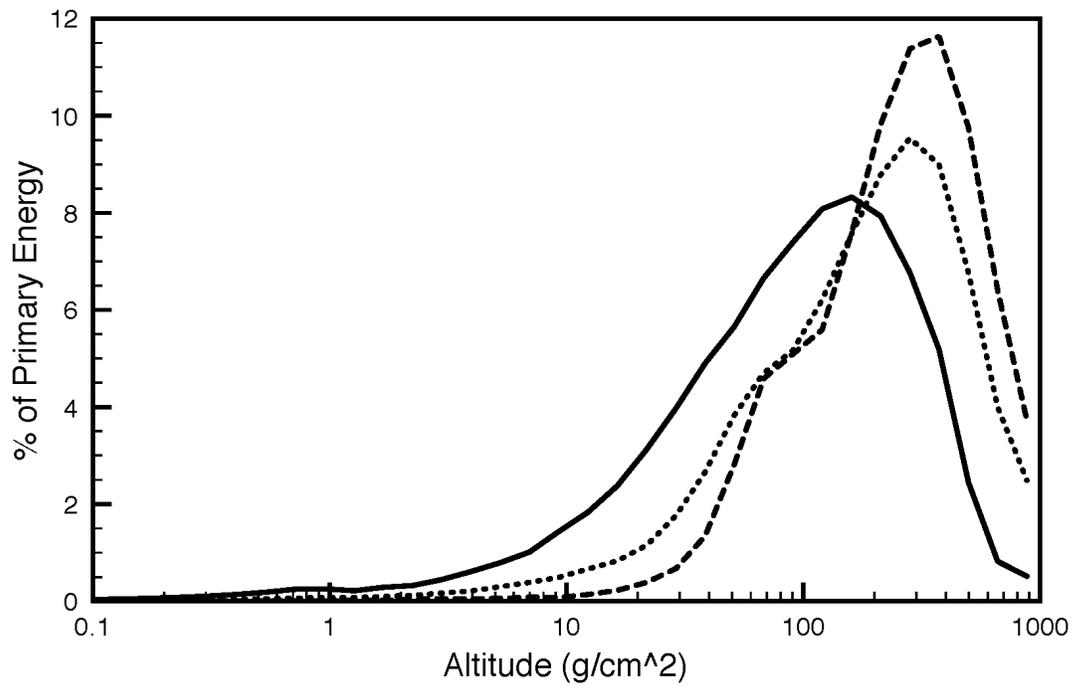

Figure 2b: Atmospheric energy deposition per log column density is plotted as percentage of primary energy, averaged over a hemisphere, for ensembles of 100 GeV (solid), 10 TeV (dotted), and 1 PeV (dashed) primaries. Deposition depth varies with incident energy.

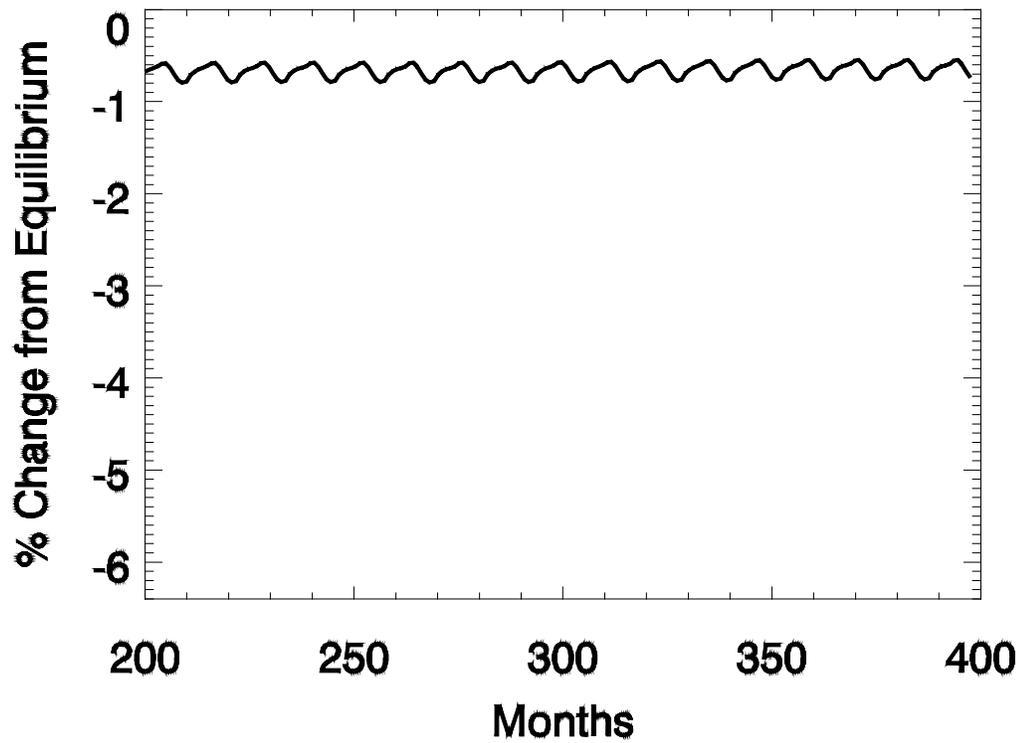

Figure 3a: Percent changes in the atmospheric $O_3$ column density as a function of time after initial transients have stabilized, averaged on the entire surface of the Earth. Oscillations are due to annual and 11-year solar cycles. In this, Case 1, the changes are too small to have any significant biological impact. Oscillations are due to the annual variation in the incidence angle of solar radiation.

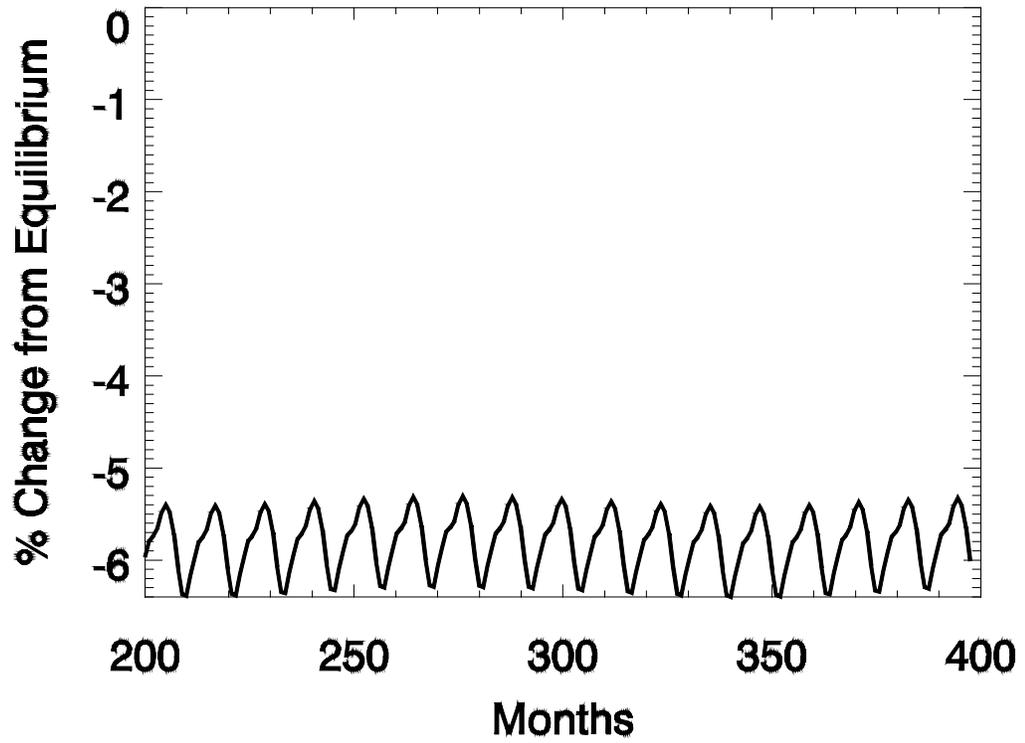

Figure 3b: As in Figure 3a, but for Case 2. Here, the size of the effect is about twice the present anthropogenic $O_3$ depletion, and so should have major impact given the several My duration of the effect.

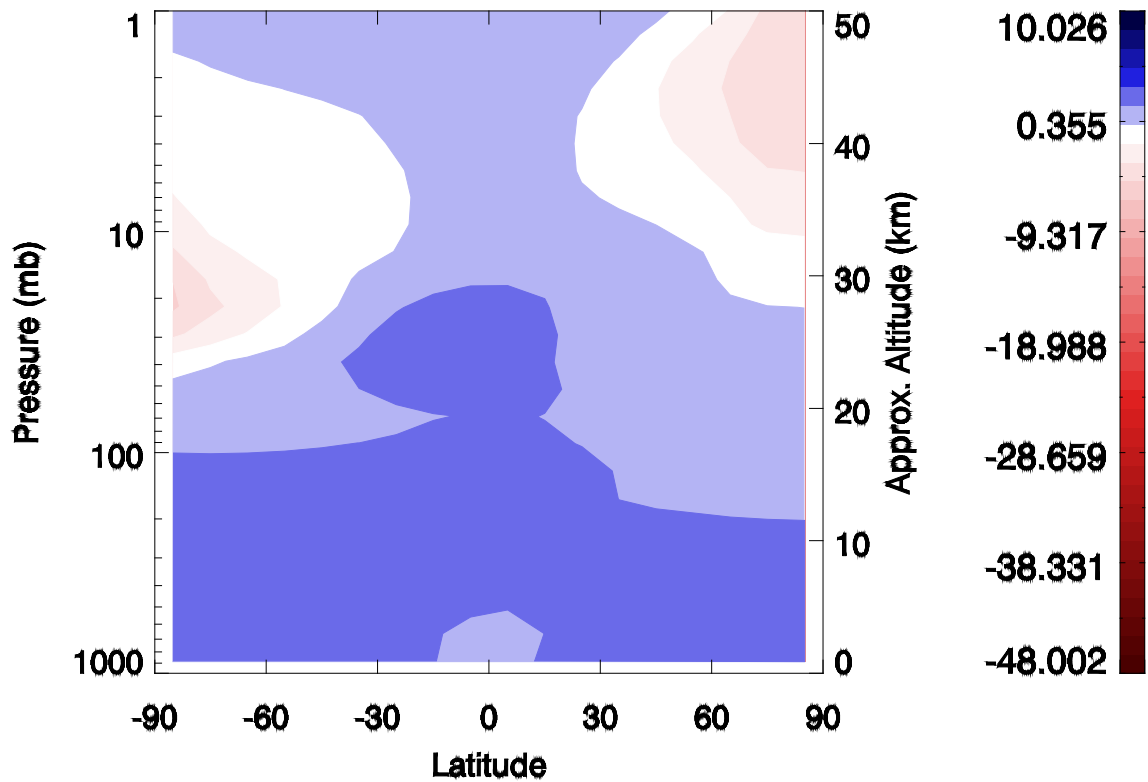

Figure 4a: A latitude, altitude plot of the percentage change in $O_3$ volume number density during March in Case 1 due to the enhanced CR background. Note that stratospheric depletion is partially compensated by production at lower altitudes. The production comes both from CR as well as increased UVB leaking through the $O_3$–depleted upper atmosphere. At high altitude near the poles is depletion, elsewhere is synthesis. But since there is little ozone near the ground in the unperturbed state, the synthesis is small; the net effect over the globe is depletion.

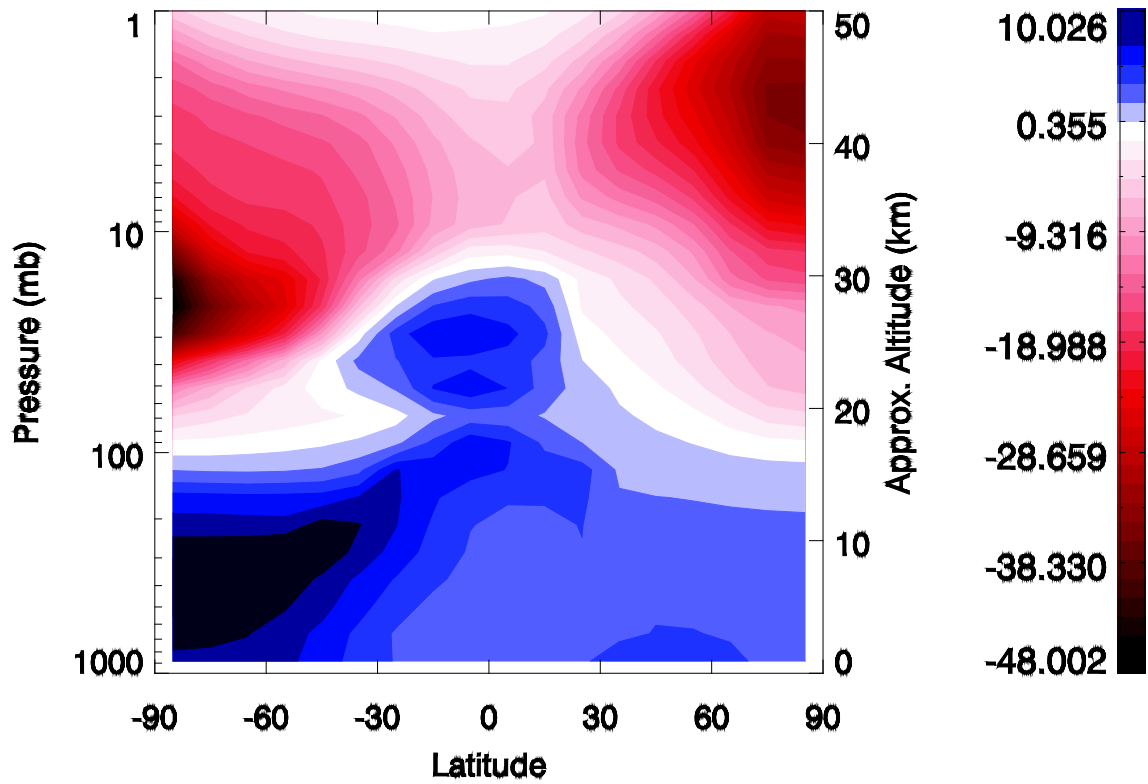

Figure 4b: As in Figure 4a, but for Case 2. In this case the depletion is much stronger, as discussed in the text.

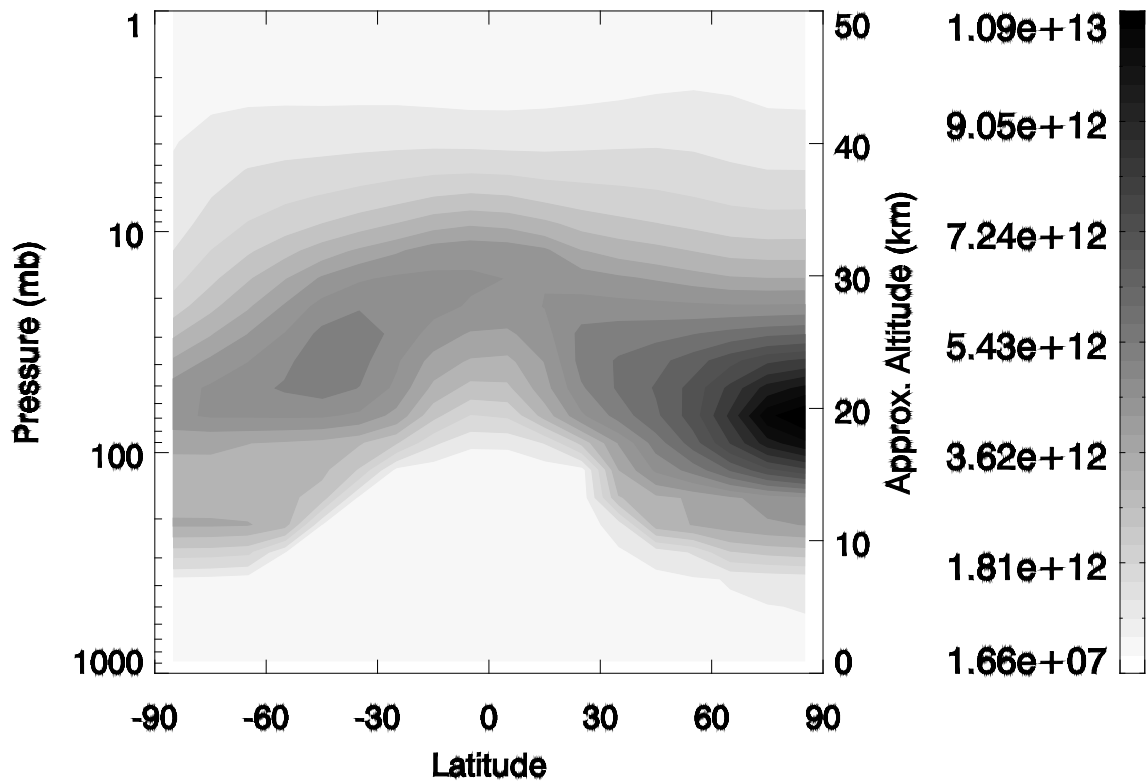

Figure 5: A latitude, altitude plot of the O$_3$ density in the unperturbed atmosphere, in March.